\pdfoutput=1
\documentclass[preprint,pra,floatfix,amssymb]{revtex4}

\newcommand{\beq}{\begin{eqnarray}}
\newcommand{\eeq}{\end{eqnarray}}
\newcommand{\half}{$\frac{1}{2}$}

\newtheorem{thm}{Theorem}
\newtheorem{prop}[thm]{Proposition}
\newtheorem{lem}[thm]{Lemma}

\newtheorem{dfn}[thm]{Definition}

\usepackage[cmex10]{amsmath}
\usepackage[mathscr]{eucal}
\usepackage{amssymb}
\usepackage{graphicx}
\usepackage{array}

\begin{document}

\title{Finite Controllability of Infinite-Dimensional Quantum Systems}
\author
{Anthony M. Bloch\thanks{A.M.B. is at the Department of
Mathematics, The University of Michigan, Ann Arbor, MI 48109.
Research partially supported by NSF grants DMS
0305837 and  DMS 604307.}, Roger W. Brockett\thanks{R.W.B is at the Department of
Engineering and Applied Sciences, Harvard University, Cambridge,
MA 02138. This work was supported in part by  the National Science
Foundation under Yale prime CCR 9980058, the US Army Research
Office under Boston University prime GC169369 NGD, the National
Science Foundation grant number EIA 0218411, and DARPA under
Stanford Prime PY-1606 and NSF 0332335}, Chitra Rangan\thanks{C.R. is
at the Department of Physics, University of Windsor, ON, N9B 3P4.
Canada. This work was supported during 2003 by FOCUS - NSF Physics Frontiers Center at the University of Michigan, and subsequently by the Natural Sciences and Engineering Research Council of Canada.} }

\begin{abstract}
Quantum phenomena of interest in connection with applications to
computation and communication almost always  involve generating specific
transfers between
eigenstates, and their linear superpositions. For some quantum systems, 
such as spin systems, the quantum evolution equation
(the Schr\"{o}dinger equation) is finite-dimensional and old
results on controllability of systems defined on on Lie groups and
quotient spaces
provide most of what is needed insofar as
controllability of non-dissipative systems is concerned. However,
in an infinite-dimensional setting, controlling the evolution of quantum
systems often presents
difficulties, both conceptual and technical.  In this paper
we present a systematic approach to a class of such problems for which
it is possible to avoid some of the technical issues.  In particular, we
analyze controllability for infinite-dimensional bilinear systems under
assumptions that make controllability  possible using  trajectories
lying in a nested family of  pre-defined subspaces.  This result,  which
we call the  Finite Controllability Theorem,  provides a set of
sufficient conditions for controllability in an infinite-dimensional
setting.  We consider specific physical systems that are of interest for quantum
computing, and provide insights into the types of quantum operations (gates) that may be developed.
\end{abstract}
\maketitle

\section{Introduction}

Over the last three decades, there has been a steady stream of
papers in the physics/chemistry literature
~\cite{HuangJMP1983,ShapiroJCP1986,TannorRice1,TannorRice2,PiercePRA88} that
describe new experiments in atomic and molecular science, and new
ways of thinking in which control-theoretic ideas are of central
importance. More recently, the driving force has been the desire
to manipulate quantum states in ways that would make possible
quantum computation or quantum communication, see for
example~\cite{RanganPRA2001,KosloffPRL02,WhaleyPRA02,KhanejaPRA00,IvanovPRL03,JacobsPRL2007}.
Phenomena involving the interaction between electromagnetic
radiation (light) and matter (e.g. ions, spin states, etc.) are
especially interesting because they are possible paradigms of
future quantum computing devices~\cite{QIC2001}. Many of the
exciting ideas are related to the control of these systems (see
for example, work on control of trapped-ion quantum
states~\cite{LawEberly,KneerLaw,BenKish,BRBCDC2003,RanganPRL2004,RanganJMP2005}).
Some infinite-dimensional systems can be made to be effectively
finite-dimensional by either bandwidth limits imposed by the
control fields~\cite{RanganPRA2001}, or by turning off specific
transitions in order to truncate the Hilbert
space~\cite{RanganPRL2004}, and the controllability of such
systems can be analyzed using finite-dimensional
methods \cite{Brockett1972,Brockett1973,RamakrishnaPRA1995,Albertini}.

We are interested in the quantum systems that are modelled as
finite-dimensional for quantum computing purposes, when in fact
they are infinite-dimensional.  The well-known paper by Huang,
Tarn and Clark~\cite{HuangJMP1983} seems to be overly pessimistic
with repect to the control of infinite quantum systems by asserting that ``using piecewise-constant controls, global controllability cannot be achieved with a finite number of
operations''.  In 2000, Zuazua summarized the field aptly thus: "From a mathematical point of view this [when the finite-dimensional system approaches the infinite-dimensional Schr\"odinger equation] is a very challenging (and very likely difficult) open problem in this area"~\cite{Z2000}.  Recently, Turinici and Rabitz~\cite{TuriniciJPA2003} adapted the classical Ball-Marsden-Slemrod results~\cite{BMS1982} to the quantum setting and showed that exact controllability does not hold in infinite-dimensional quantum systems (see also \cite{Beauchard2005}).  Along with an excellent review, Illner, Lange and Teismann~\cite{TeismannESAIM2006} show that the Hartree equation that is well-known in quantum chemistry with bilinear control is not controllable in finite or infinite time.  In spite of these negative results, most quantum computing systems that are indeed
infinite-dimensional have shown themselves to be remarkably
amenable to the production of a variety of states.  Is it possible
then to make a statement regarding the reachable set of states in
such systems?  

There has been recent progress in attacking this
problem, and a few positive results.  Ref.~\cite{TuriniciCDC2000} showed that for an infinite-dimensional quantum control system with bounded control operators a quantum state
could be steered only within a dense subspace of the relevant Hilbert space.
In 2003~\cite{BRBCDC2003}, we showed that
one can reach any finite linear superposition of states
in a two-level system coupled to a harmonic
oscillator by the alternate application of control
fields even when one of the control operators is unbounded. 
Ref.~\cite{ShapiroJCP2004} presents an ingenious method of
creating arbitrary finite superpositions of rotational states of
molecules by ensuring the trajectories lie within SU(N)[z,z-1]
loop group of N-periodic matrices. Ref.~\cite{KarwowskiPLA2004} presents a scheme that is well-known in atomic physics - when an infinite-dimensional quantum system has unequally spaced bound state energy levels, single resonant fields can be used to transfer population (albeit very slowly).  Ref.~\cite{LanJMP2005}
provides a prescription for determining whether there exists a
submanifold that is strongly analytic controllable, but this
prescription does not identify the submanifolds.  Ref.~\cite{Adami2005} presents an adiabatic method of controlling a sequentially connected system in which the transition couplings get weaker as one moves away from the ground state.  Ref.~\cite{WuPRA2006} presents an algebraic framework to determine the nature of controllability of some infinite-dimensional quantum systems, specifically those with continuous spectra.    Ref.~\cite{Boscain2009} proves approximate controllability for bound states of quantum systems that are not equally spaced in energy even when the control matrices are unbounded.

We begin this paper in Section \ref{finite} by discussing the difficulties associated with applying traditional methods of
analysis to infinite-dimensional control systems.  We also
discuss a specific class of infinite-dimensional systems --- one in which
it is desirable to steer a finite superposition of eigenstates to another finite superposition of eigenstates.  We present a general theorem on Finite Controllability of quantum systems that exhibits the conditions needed for such transfers.  For the specific quantum-computing models of interest, the
trajectories are constrained to lie within subspaces.  Section \ref{examples}
presents examples of quantum-computing systems such as a model
of a trapped-ion qubit and trapped-electron qubits. We show how our theorem can used to be
prove finite controllability in the former case. We also discuss other similar
systems where the theorem can and cannot be applied.

\section{Infinite-Dimensional Controllability}
\label{finite}
Controllability results for infinite-dimensional systems are
seldom just straightforward extensions of the finite-dimensional
ones, and in particular this is true for bilinear systems.
Recently, there has been significant interest in the class of
bilinear systems because of their relevance to quantum
control.  In the following subsection we illustrate the limitations of
applying the tools of finite-dimensional systems analysis to certain
classes of infinite-dimensional systems.  

\subsection{Limitations of Lie Algebraic analysis}
Examining the Lie algebraic structure often gives us insights
into the controllability of a quantum system, but in the case of
infinite-dimensional systems, this insight is limited.  In the
well-known example of a resonantly-driven quantum harmonic
oscillator (see e.g. ~\cite{Schiffbook}, \cite{BRBCDC2003} and \cite{Rouchon},
the evolution is given by
\beq
\frac{\partial \psi }{\partial t} = \left( \omega\frac{i}{2}\left(\frac{\partial
^2}{\partial x^2} -x^2\right) -i u(t)x \right) \psi. \eeq

Here, the bilinear control term $u(t)x$ arises
because of the dipole interaction between the field and harmonic
oscillator. The two operators of interest, $ A= \frac{i}{2}\left(
\frac{\partial ^2 \psi }{\partial x^2}-x^2 \right)$ and $B=-ix$
generate a Lie algebra of skew-hermitian operators that is just
four-dimensional. Thus, we expect that the control of this system
will be limited.  For example, it is well-known that it is not
possible to transfer the number state $x(0)=|0\rangle$ to
$x(T)=|n\rangle$ for $n>0$~\cite{ShoreBook} using this control. We return to this example in Section \ref{examples}.

However, even when we encounter infinite-dimensional systems for which
the Lie algebra also is infinite-dimensional (such as in Ref.\cite{LloydPRL1999,BRBCDC2003}), the statements one can make about the controllability are also limited.  More work is
required to say with precision exactly what the reachable states
are.  This issue is  illustrated in Section \ref{examples}.


In computing the span of the Lie algebra, it is necessary, of course that the domain of the operators involved be such that the bracket operations are allowable.  This is the case in this paper, but we do not discuss these technical details here.

\subsection{Finite  Controllability}
In this section, we prove an elementary but useful  theorem about
controllability
on finite-dimensional subspaces of a complex Hilbert space. It is the
basis for the applications we will present in the next section.  

\begin{dfn}
We will say that a finite-dimensional system evolving
in the space of complex $n$-vectors $x$, 
$$ \dot{x} = \sum_i u_iG_ix $$
with skew-Hermitian operators $G_i$, is {\em  unit vector controllable} if any unit length
vector $x_0$ can be steered to any second  unit length vector
$x_f$ in finite time.
\end{dfn}

\begin{thm}[Finitely Controllable  Infinite Dimensional Systems]
\label{infinite}
Consider a complex Hilbert space $\mathcal X$ together
with a nested  set of  finite-dimensional subpaces $\mathcal H = \{
\mathcal H _1\subset  \mathcal H _2 \subset \mathcal H _3 ... \} $.
Consider
$$ \dot{x} = (\sum _{i=1}^m u_i G_i)x.$$
 Assume that $\mathcal H
_1$ is an invariant subspace for a subset $\mathcal G _1$ of the set  $\{ G_i \}$
and that the system is unit vector controllable on $\mathcal H _1$ using
only  this subset of the $G_i$.    If for each $\mathcal H _\alpha \; \alpha \ne 1 $  there is a
subset  $\mathcal G _\alpha $  of  $\{ G_i \} $ that  leaves
$\mathcal H _\alpha $ invariant and if  for any unit vector in
$\mathcal H _\alpha $ the orbit generated by exp $\left( \mathcal G
_\alpha  \right) $  contains a point in one of the lower dimensional
subspaces   $\mathcal H _\beta$ then any unit vector  in any of the
$\mathcal H _i$   can be steered to any other unit vector in any other
$\mathcal H _j$ using a finite number of piecewise constant controls.

\noindent {\bf Remark:}
Given a system and a nested set of finite dimensional subspaces it will be said
to be  {\em
finitely  controllable} if it can be transfered from any point in one of the subpaces to any other point in that subspace with  a trajectory lying entirely within the subspace. 
\end{thm}

\noindent {\textbf Proof:}
Suppose that  $x(0)\in \mathcal H _\alpha $ and $x(T) \in
\mathcal  H _\beta $  are unit vectors representing the initial value of
$x$ and the desired final value, respectively.  Then by assumption
either there exists  a subset of the $\{ G_i \}$ that leaves $\mathcal
H _\alpha $ invariant and steers $x(0)$ to a point in some  $\mathcal H _\beta
$ with the dimension of $\mathcal H _\beta $ being strictly less than
that of $\mathcal H _\alpha $ or else $x(0)\in \mathcal H _1$.   A finite
induction on the index of the set $\mathcal H _i $  then shows that the state can be
steered to a unit vector in the subspace $\mathcal H _1$,  which is a controllable subspace.  To finish the proof,
observe that if one can reach $x_a\in \mathcal H $ from $x_b \in
\mathcal H $ then the  standard time reversal argument using the fact that the $G_i$ are skew-Hermetian shows that it is
possible to reach $x_b\in \mathcal H $ from $x_a \in \mathcal H $. Thus
a second application of the steps given above implies that it is
possible to reach $x(T)\in \mathcal H _\beta $. 

{\it Discussion and example:}
\label{discussion}
Various technical issues that arise in infinite-dimensions illustrate the significance of this theorem.  All
finite-dimensional subspaces are closed in the usual topology but
infinite-dimensional subspaces need not be and there are a number
of issues involving infinite-dimensional controllability that can
be clarified by giving more attention to the distinction.  Let $l_2$ denote the Hilbert space of infinite vectors whose entries
are square summable which corresponds to the entire
state space of our quantum system. Let $l_0$ denote the subspace consisting of
those elements with only a finite number of nonzero entries, which
corresponds in our setting to a finite superposition of states. Of
course, this subspace is not closed.

We recall now some general ideas from operator theory \cite{Dunford} and discuss
their relationship to our theorem.
\begin{enumerate}
\item If $e^A$ and $e^B$ are bounded operators mapping a
Hilbert space $\mathcal H$ into itself and if $\mathcal V$ is an
invariant subspace for $e^A$ and $e^B$, then $\mathcal V $ is
invariant for the product $e^Ae^B$.  

In our setting $e^{B_c}$ and $e^{B_r}$
are unitary and thus bounded.
\item If $A$ is the infinitesimal generator of a semigroup
mapping a Hilbert space $\mathcal H$ into itself and if $\mathcal
V$ is a closed, nontrivial, invariant subspace for $A$, then $\mathcal V $ is
invariant subspace for $e^{At}$.

Consider for example
an operator $A$ on $l_2$ that maps the $i^{th}$ unit
basis vector $e_i$ into $e_{i+1}$ for all $i= 1,2,...$. This
operator clearly sends $l_0$ into itself but $e^{At}$ for $t\ne 0$
sends $e_1$ into the element $\sum_i e_it^i/i!$ which is not in
$l_0$.
\item If $A$ and $B$ are infinitesimal generators of a
semigroup mapping a Hilbert space $\mathcal H$ into itself and if
$\mathcal V$ is a closed, invariant subspace for $A$ and $B$, then
$\mathcal V $ is invariant subspace for $e^{(A+B)t}$.

Again, the assumption that $\mathcal V$ is closed
cannot be dispensed with. 
For example, consider a system $\dot{x}=uAx+vBx$ with  $A$ being block
diagonal with two-by-two blocks and $B$ being basically block diagonal
with  two-by-two blocks except for top and left borders where  the necessary
one-by-one, two-by one  and one-by two matrices pad an otherwise block
diagonal matrix.  Specifically, $A$ and $B$ are given by
\beq A & = \left[ \begin{array}{ccccc} B_0 & 0_{22} & 0_{22} & 0_{22} & ...
\\ 0_{22} & B_0 & 0_{22} & 0_{22} & ... \\ 0_{22} & 0_{22} & B_0 & 0_{22} &
... \\ 0_{22} & 0_{22} & 0_{22} & B_0 & ... \\ ... & ... & ... & ...
& ... \end{array} \right] \;\;;\;\;
\\B & = \left[ \begin{array}{ccccc}
0_{11} & 0_{12} & 0_{12} & 0_{12}& ... \\ 0_{21} & B_0 & 0_{22}
&0_{22}&... \\  0_{21} & 0_{22} & B_0 & 0_{22} & ...\\ 0_{21} & 0_{22} &  0_{22}  
&  B_0 & ... \\ ... & ... &... &... & ... \end{array}
\right], \eeq
with $0_{ij}$ denoting an $i$ by $j$ matrix of zeros and
\beq  B_0 = \left[ \begin{array}{cc} 0 &1 \\ -1 & 0 \end{array} \right]. \eeq
The operators $A$ and $B$ leave $l_0$ invariant but the exponential of their sum does not.
\end{enumerate}

One can provide a  direct  argument not involving
Lie theoretic techniques that shows that any unit vector can be
transferred to $e_1$.  The idea, motivated by the analysis in Ref.~\cite{LawEberly}, is to alternate the use of $A$ and $B$  
and reason that if we start with a vector with $x_n\ne 0$ and $x_{m>n}=0$ we can use a
control with $v=0$ (or $u=0$ depending on whether $n$ is even or odd) to reduce the vector to one for which $x_n$ is zero,
then use a control with $u=0$ to reduce the vector to one with
$x_{n-1}=0$ without changing $x_n$, etc.  This argument depends on
observing that  
\beq e^{Au}=&\left[ \begin{array}{ccccc} e^{B_0u} & 0_{22} & 0_{22} & 0_{22} &
... \\ 0_{22} & e^{B_0u} & 0_{22} & 0_{22}& ... \\ 0_{22} & 0_{22} &
e^{B_0u} & 0_{22} & ...  \\ 0_{22} &
0_{22} & 0_{22} & e^{B_0u} & ...  \\ ... & ... & ... & ... & ...\end{array} \right] \;\;;\;\;\\ 
e^{Bv}=&
\left[ \begin{array}{ccccc} 1 & 0_{12} & 0_{12} & 0_{12}& ... \\ 0_{21}
& e^{B_0v} & 0_{22} &0_{22}&... \\ 0_{21} &  0_{22}  &  e^{B_0v}  & 0_{22} &... \\ 0_{21} & 0_{22} &0_{22} &e^{B_0v}  &
...  \\  ... & ... & ... & ... & ...\end{array} \right]  \eeq
from which we see that  $e^{Au}$ simply rotates the entries in $x$
pairwise, i.e., $x_1,x_2$; $x_3,x_4$; etc.  so that by attending only to
the subspace defined by $x_n$ and $x_{n-1}$ one can make $x_n=0$. Having
done so, we can use $e^{Bv}$ to rotate in $x_{n-1},x_{n-2}$ space so as
to reduce $x_{n-1}=0$ with out changing $x_{m>n-1}$, etc.      This shows that the  Lie group  generated by $e^{Au}$ and $e^{Bv}$ acts
transitively on the unit vectors and thus identifies the Lie algebra to
within one of a few possibilities. (See Ref.~\cite{Brockett1973}).  

\section{Physical Systems}
\label{examples}
In this section, we apply the Finite Controllability
Theorem to determine the reachable set of states of four
infinite-dimensional quantum systems, namely, the quantum harmonic
oscillator, a spin-half system coupled to a harmonic oscillator
(model of a trapped-ion qubit), an N-level atom coupled to harmonic oscillator, and a spin-half system coupled to
two harmonic oscillators (model of a trapped electron qubit). 
Note that in all sections below we use atomic/scaled/dimensionless units so that the evolution equations do not explicitly contain Planck's constant $\hbar$, the charge of the electron $e$ or the mass of the electron $m_e$.

\subsection{System 1: Quantum harmonic oscillator}
We discuss this well-known system first. The controllability algebra is finite-dimensional and, in
particular, the system  does not satisfy the conditions needed for the application of the Finite Rank
Controllability Theorem.  The discussion, however, is key to setting up the formalism used in subsequent examples that are finitely controllable.

\begin{figure}[htbp]
\centering
\rotatebox{0}{\includegraphics[width=3in]{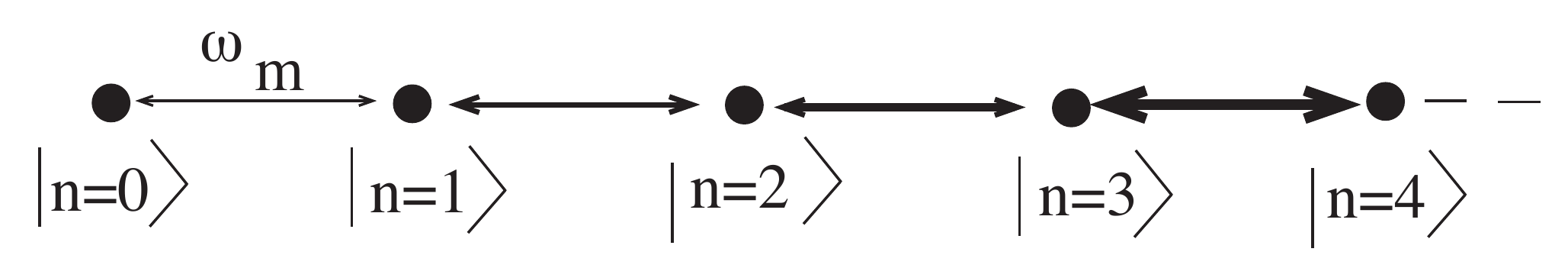}}
\caption{Graphical
representation of the quantum harmonic oscillator driven by a
sinusoidal resonant field.  Note that while the strengths of the transition
couplings increase as the square root of the quantum number $n$ 
as shown by the boldness of the connections between energy levels, the
transition frequency between each level is the same. It is this latter
property that leads to a lack of controllability. \label{fig:HO}}
\end{figure}

The problem of controlling the harmonic oscillator has been
discussed many times (see e.g. \cite{HOcontrol} and other references as
discussed above).
If the control is a sinusoidal resonant driving field (of frequency equal to the harmonic oscillator frequency $\omega_m$) as shown in the transfer graph
Fig.~\ref{fig:HO}, then the evolution is via \beq \frac{\partial
\psi }{\partial t} = \left( \omega_m\frac{i}{2}\left(\frac{\partial
^2}{\partial x^2} -x^2\right) -i u(t)x \right) \psi . \eeq
Here, the control term $u(t)x$ arises because of the dipole
interaction between the field and harmonic oscillator. The
operators of interest are $ A= \frac{i}{2}\left( \frac{\partial
^2 }{\partial x^2}-x^2 \right)$ and $B=-ix$.  $A$ and $B$ generate a Lie algebra of
skew-hermitian operators that is just four-dimensional ($C=[A,B] =\frac{\partial }{\partial x}$, $D = [B,C] =iI$, where $I$ is the identity operator). 
This in itself tells us that the
resonantly driven harmonic oscillator is not controllable.

As is well-known, the spectrum of $A$ is discrete. 
If we
describe the evolution in terms of an eigenfunction expansion,
with the basis being $|n\rangle$'s, the eigenfunctions of $\partial ^2/\partial
x^2 - x^2$, then the evolution is via \beq \dot{x}_n & = &  -i\omega_m (n+\frac{1}{2})x_n \\ \nonumber
&   & - u(t)\frac{i}{\sqrt 2}(\sqrt{n-1}x_{n-1} -\sqrt{n}x_{n+1}) .\eeq
Although the eigenstates of the harmonic oscillator can be
written as an infinite set of nested finite subspaces, it is seen
that the operator $B$ connects space $\mathcal H_i$ to \emph{both}
$\mathcal H_{i-1}$ and $\mathcal H_{i+1}$.  Thus finite superpositions of
eigenstates may not be reached by resonantly driving the harmonic
oscillator, consistent with the fact that the requirements of the 
Finite Controllability Theorem are not met. Physically, this is due to the degeneracy of spacings between the eigenstates and
the fact that the control vector field simultaneously
illuminates all states.

However, the system is in fact partially controllable.
One can be quite explicit about the reachable wave
functions under the application of a control~\cite{ShoreBook}. It is interesting that the finite-dimensionality of the algebra that makes the system uncontrollable is also the property that makes it possible to derive an analytic expression for the unitary propagator~\cite{Rau}.  For example, if we
agree to call any function of $t$ and $x$ which has the form \beq
\psi (0,x) = k(t)e^{i(a(t)x^2 + b(t)x)} \eeq {\em i-gaussian },
then if the initial value of $\psi $ is i-gaussian the solution
will be i-gaussian for all time, regardless of the choice of
$u(t)$. (This is directly analogous to the fact that the solution
of the conditional density equation of estimation theory remains
gaussian if it has a gaussian initial condition.) An example of such
a function is the well-known coherent state~\cite{GlauberPRL1963}
\begin{equation}|\alpha\rangle  = \exp(-|\alpha|^2/2)\sum_{n=0}^{\infty}
\frac{\alpha^n}{n!}|n\rangle.
\end{equation}
We see that it is not possible to transfer $x(0)=|\alpha\rangle$ to
$x(T)=|i\rangle$ for $i>0$ because $|i\neq 0\rangle$ is not i-gaussian,
but $|\alpha\rangle$ is, and is in fact a coherent state.

\subsection{System 2: Spin-half particle in a quadratic potential}

In contrast to the harmonic oscillator,
the model of a spin-half particle coupled to a harmonic oscillator with
suitable controls turns out to be finitely controllable.
This model is a good representation of an ion with two essential internal
states trapped in a quadratic potential.  We show below
that this system satisfies the conditions of the Finite Controllability Theorem of Section \ref{infinite}.
Moreover, one can also provide
an algorithm for explicit control. The spin-\half  ~model
represents a two-level atomic ion with an energy splitting $\hbar
\omega_0$, where the frequency $\omega_0/2\pi$ is in the several
GHz range. The atomic levels are coupled to the motion of the ion
in a harmonic trap~\cite{WinelandNISTreport1998}. These quantized
vibrational energy levels are separated by a frequency
$\omega_m/2\pi$ in the MHz range.

In a frequently cited paper, Law and Eberly~\cite{LawEberly}
showed that when properly interpreted, this system has interesting
controllability properties, quite different from the properties of
the harmonic oscillator alone. In fact, by coupling the harmonic
oscillator with a two-level system it is possible to arrive at a
system which is much more controllable than the harmonic
oscillator. At an intuitive level, this can be seen simply as a
consequence of the fact that the addition of a spin degree of
freedom breaks the infinite degeneracy associated with the
harmonic oscillator and allows the system to resonate with more
than one frequency. This allows the transfer of population from
any eigenstate to any other eigenstate by sequentially applying
the two frequencies.  We now analyze this system from a
controllability viewpoint.

An eigenstate of the spin-half system coupled to a quantum harmonic oscillator is denoted by $|S, n \rangle $, where the first index refers to the ``spin'' state of the system, and the second index is the number state of the harmonic oscillator.  An applied field causes transitions between the eigenstates of the coupled spin-oscillator
system.   A monochromatic field of angular frequency $\omega=\omega_0$ causes
resonant transitions between states $|\downarrow,n\rangle$ and
$|\uparrow,n\rangle$ (carrier or spin-flip transitions).  A monochromatic field of angular frequency $\omega=\omega_0-\omega_m$  causes resonant transitions between
states $|\downarrow,n\rangle$ and $|\uparrow,n-1\rangle$, i.e., produces so
 called red sideband (that is with angular frequency $\omega=\omega_0-\omega_m$)
 transitions.

\begin{figure}[h!]
\centering
\rotatebox{0}{\includegraphics[width=3in]{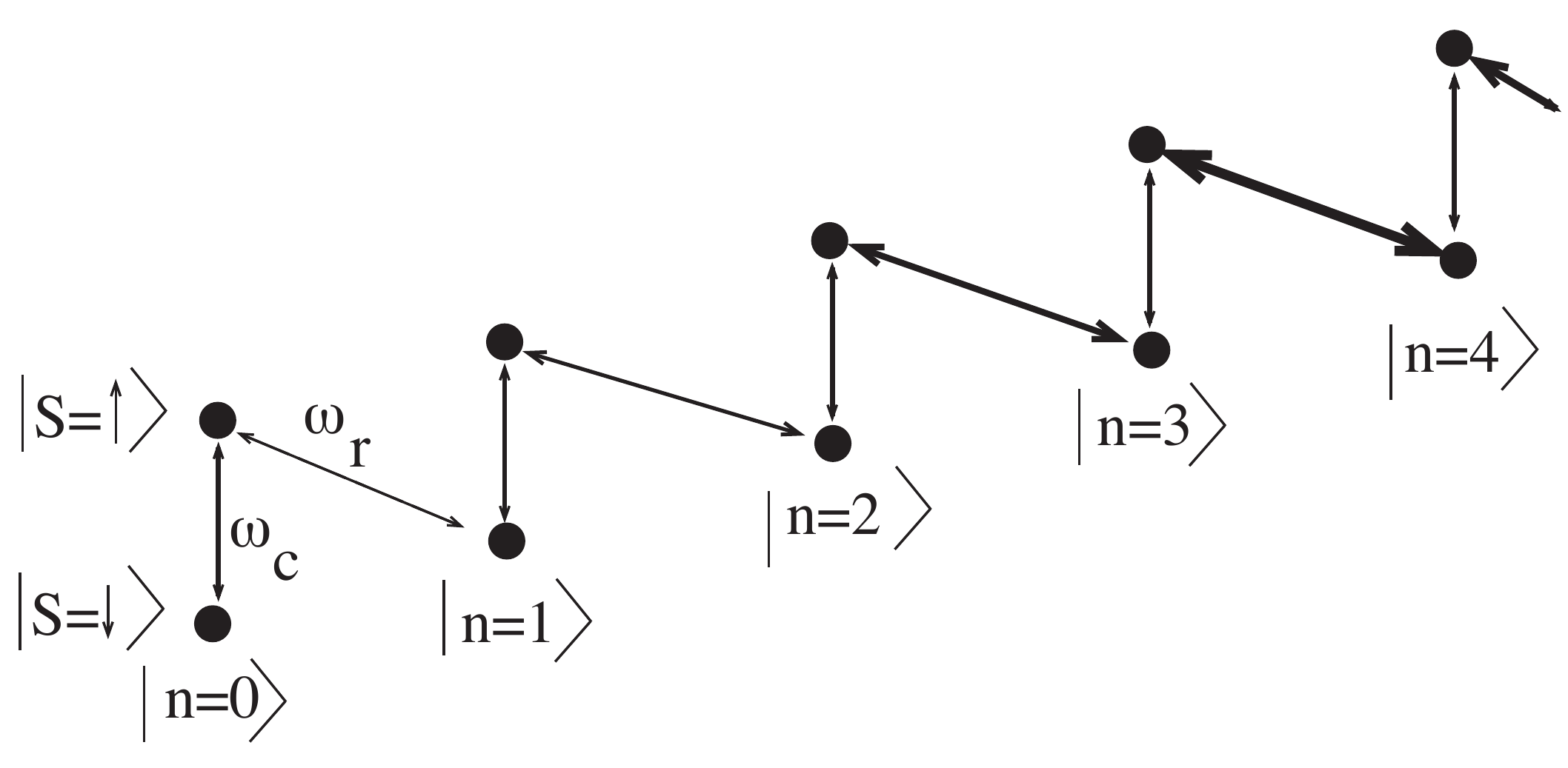}}
\caption{Graphical representation of the coupled spin-half quantum
harmonic oscillator system driven by sinusoidal resonant fields of angular frequency
$\omega_c$ and $\omega_r$ as shown. When $\eta \ll 1$, the strengths of the
$\omega_c$ transition couplings are independent of the harmonic
oscillator quantum number $n$, whereas the strengths of the
$\omega_r$ transition couplings increase as the square root of
$n$ as shown by the boldness of the coupling lines.\label{fig:spin2HO}  Note
that there is no direct coupling between two consecutive oscillator states with fixed spin. }
\end{figure}

These transitions
are graphically depicted in Fig.~\ref{fig:spin2HO} with the
thickness of the edges qualitatively representing the strength of
the coupling between the states.  As pointed out in
Ref.~\cite{RanganJMP2005}, when both fields (carrier and red sideband) are applied {\it
simultaneously}, the eigenstates of the system are sequentially
connected.   Therefore, we look at the trapped-ion model controlled only by these two fields.

Now we write the evolution equation of the spin-half coupled to harmonic oscillator driven by two fields that drive the carrier and red sideband transitions.  The amplitudes corresponding to the fields that cause the carrier and red transitions are dubbed $E_c$ and $E_r$
respectively.  As detailed in Appendix~\ref{appspinHO}, in the interaction picture and in the energy eigenbasis, the evolution equation is written as
\begin{eqnarray}
\dot{Y} & = & (u(t) B_c + v(t) B_r)Y. \label{controlsystem}
\end{eqnarray}
The controls $u(t)$ and $v(t)$ are related to the applied fields via the equations
\begin{eqnarray}
u(t) & = & c_1 E_c(t) = 0.25\mu \exp(-\eta^2/2) E_c(t), \\
v(t) & = & c_2 E_r(t) = 0.25 \eta \mu \exp(-\eta^2/2) E_r(t).
\end{eqnarray}
Here  $\eta$, the so-called Lamb-Dicke parameter, is the 
product of $k$, wave vector of the light, and $x_0$, the
amplitude of the zero-point motion of the particle in the harmonic
potential (or the spatial extent of the ground state harmonic
oscillator wave function).    By ordering the eigenstates as
$|\uparrow,0\rangle,\ |\uparrow,1\rangle,\ \ldots,
|\downarrow,0\rangle,\ |\downarrow,1\rangle,\ \ldots$, the control matrices are written as
\begin{eqnarray}
B_c &=& \left(\begin{array}{c|c}
0 & i L_0 \\
\hline
i L_0^T & 0\\
\end{array} \right).\\
B_r &=& \left(\begin{array}{c|c}
0 & L_1 \\
\hline
-L_1^T & 0\\
\end{array} \right).
\eeq
The upper-triangular matrices $L_0$ and $L_1$ are defined as \beq
L_0 & = & \left(\begin{array}{cccc}
L_0(\eta^2) & 0 & 0 & \ldots \\
0 & L_1(\eta^2) & 0 &  \ldots \\
0 & 0 & L_2(\eta^2) & \ddots \\
\vdots & \vdots & \vdots & \ddots \\
\end{array} \right). \label{eq:L0}\\
L_1 & = & \left(\begin{array}{cccc}
0 & L_0^{(1)}(\eta^2) & 0 & \ldots \\
0 & 0 & L_1^{(1)}(\eta^2) & \ldots\\
0 & 0 & 0 & \ddots \\
\vdots & \vdots & \vdots & \ddots \\
\end{array} \right). \label{eq:L1}
\end{eqnarray}

This structure of the control Hamiltonian precludes the use of adiabatic methods such as those used in  Ref.~\cite{Adami2005}.  Unlike the situation encountered in the analysis of the quantum harmonic oscillator
algebra, here the formal commutation of the operators $B_c$ and
$B_r$ does not lead to a finite-dimensional algebra, suggesting
that the model with spin is much more controllable. This is the
case, as will be explored in the next subsection.

\subsubsection{Controllability: Lie Algebra}

It is interesting to compare our analysis with the formal calculations
suggested by Lie theory. The first thing to do is to determine the formal
structure of the Lie algebra, which we now consider. 

The control of the trapped-ion system is often studied in two different
limiting cases  - one in
which the extent of zero-point motion of the spin-half particle in
the harmonic potential $x_0$ is much smaller than the wavelength
of the applied light $2\pi/k$, i.e., $\eta \ll 1$ (the Lamb-Dicke limit), and the other in which $\eta \simeq
1$ (beyond the Lamb-Dicke limit).  The case in which $\eta \simeq 1$ is more general than the case of the Lamb-Dicke limit, but requires a more sophisticated analysis.  We study initially the Lamb-Dicke limit in which the Lamb-Dicke parameter $\eta \ll 1$.  The terms in equations (\ref{eq:L0})
and (\ref{eq:L1}) are expanded to first order in $\eta$. The
control Hamiltonians can then be expressed in operator form as
\beq B_c  & = \left[ \begin{array}{cc} 0 & i \\
i & 0 \end{array} \right] ,
\
{\rm and} \\
B_r & =  \eta \left[ \begin{array}{cc} 0 & a \\
-a^{\dag} & 0 \end{array} \right] , \label{eq:controlops} \eeq
\noindent where $a$ and $a^{\dag}$ denote the annihilation and creation operators of the harmonic oscillator as defined in  Appendix~\ref{appspinHO}.  (Note that $B_r$ is the same Hamiltonian as obtained from the well-known
Jaynes-Cummings model~\cite{JaynesCummings} that describes the
interaction between a quantized cavity field and a two-level atom.)  

In order to compute the Lie algebra, let us consider $T$,  an
operator acting on a complex Hilbert space. We associate with $T$
a skew-hermitian operator acting on $\mathcal H \oplus \mathcal H$
defined by \beq J(T) = \left[ \begin{array}{cc} 0 & T \\
-T^{\dagger} & 0 \end{array} \right]. \eeq For convenience, let
$K(T)$ be another operator defined in a similar way as \beq K(T) =
\left[ \begin{array}{cc} T & 0 \\ 0 & -T \end{array} \right] .\eeq
Of course, $K(T)$ is skew-hermitian if and only if $T$ is.  The control operators we are interested in for the purposes of determining the structure of the Lie algebra are given by
$B_c=J(iI)$ and $B_r=\eta J(a)$. We have
\begin{lem} The Lie algebra generated by $J(iI)$ and $J(T)$
includes the operators \beq J(W^{2p}); \; p=1,2,3,\cdots\; ;
\; K(W^{2p+1}); \; p = 0,1,2,\cdots\ , \eeq where,
$W=i(T+T^{\dagger})$.
\end{lem}

\noindent {\textbf Proof:}
A calculation shows that $[J(T),J(iI)] = K(W)$ and
further, $[J(iI),K(W)]=-2iJ(W)$. We can then check that \beq
ad_{J(W)}^p (K(W))=(-2)^p\left\{\begin{array}{ll}
J(W^{p+1}),\; {\rm if \ p\ is \ odd}\\
K(W^{p+1}),\; {\rm if \ p\ is \ even}
\end{array}\right\}.
\eeq

These calculations make it clear that if the powers of $W$ are
independent then $J(iI)$ and $J(T)$ do not generate a
finite-dimensional algebra. Thus if T is nonzero only on the
diagonal immediately above the main diagonal (which is true for
the operator $a$), and if every term on this upper-diagonal is
nonzero, then the successive powers of $W$ are independent and the
algebra is infinite-dimensional.  

This is the case for the coupled
spin-half harmonic oscillator system.  Of course, this calculation
only shows that this system, unlike the harmonic oscillator,  does
not generate a finite-dimensional controllability Lie algebra.
More work is required to say with precision exactly what the
reachable states are. This is precisely the role of Theorem
\ref{infinite} which gives more specific information of 
which operators play a role in the control process. 

Note: In the case where the Lamb-Dicke limit does not apply, the
Lie algebra will still be infinite-dimensional but the terms are
more complicated.

\subsubsection{Finite controllability}

In this subsection we discuss how finite controllability works
in this infinite-dimensional setting.

From Fig.~\ref{fig:spin2HO}, it is seen that the sequentially
connected eigenstates can be looked at as an infinite set of
finite-dimensional subspaces with the ground state
$|\downarrow,0\rangle$ being equal to $\mathcal H_1$.  Further, when
operators $B_c$ and $B_r$ are applied \emph{sequentially}, each subspace
$H_i$ can be transferred to $\mathcal H_{i-1}$.  Thus the criteria for
finite controllability are met.
By sequential application of
the two operators, any finite superposition of eigenstates can be
transferred to the ground state in finite time.

The application of these statements to the spin-half in quadratic
potential example is best understood by writing the control
matrices $B_c$ and $B_r$ in a re-ordered basis as follows: The
eigenstates can be ordered as $|\uparrow,0\rangle,\
|\uparrow,1\rangle,\ \ldots, |\downarrow,0\rangle,\
|\downarrow,1\rangle,\ \ldots$. 
In the interaction picture, the Schr\"{o}dinger equation is
written as
\begin{eqnarray}
\dot{Y} & = & (u(t) B_c + v(t) B_r)Y,
\end{eqnarray}
where $u(t)$ and $v(t)$ are defined as before. Then, 
\begin{eqnarray}
B_c & = & i \left(\begin{array}{cc|cc|cc|c}
0 & L_0 & 0 & 0 & 0 & 0 & \ldots \\
L_0 & 0 & 0 & 0 & 0 & 0 & \ldots \\
\hline
0 & 0 & 0 & L_1 & 0 & 0 & \ldots \\
0 & 0 & L_1& 0 & 0 & 0 & \ldots \\
\hline
0 & 0 & 0 & 0 & 0 & L_2 & \ldots \\
0 & 0 & 0 & 0 & L_2 & 0 & \ldots \\
\hline
\vdots & \vdots & \vdots & \vdots & \vdots & \vdots & \ddots \\
\end{array} \right). \nonumber
\end{eqnarray}
\begin{eqnarray}
B_r & = & \left(\begin{array}{c|cc|cc|cc}
0 & 0 & 0 & 0 & 0 & 0 & \ldots \\
\hline
0 & 0 & L_0^{(1)} & 0 & 0 & 0 & \ldots \\
0 & -L_0^{(1)} & 0 & 0 & 0 & 0 & \ldots \\
\hline
0 & 0 & 0 & 0 & L_1^{(1)} & 0 & \ldots \\
0 & 0 & 0 & -L_1^{(1)} & 0 & 0 & \ldots \\
\hline
0 & 0 & 0 & 0 & 0 & 0 & \ldots \\
\vdots & \vdots & \vdots & \vdots & \vdots & \vdots & \ddots \\
\end{array} \right). \nonumber
\end{eqnarray}
\noindent $L_i$'s and $L_i^{(1)}$'s are Laguerre polynomials of the zeroth and first order, all with argument $\eta^2$.

One can immediately see the applicability of the general
statements made in the subsection~\ref{discussion} to this
system. The model of the trapped-ion qubit highlights the
existence of important examples for which it is desirable for the
evolution to occur on a non-closed subspace of a Hilbert space, i.
e., the space of finitely nonzero elements. In this model, this
non-closed subspace consists of vectors in the oscillator
representation with finitely many nonzero elements. The $B_c$ and
$B_r$ operators and their one parameter groups, leave invariant
the subspace $l_0$  of $l_2$ consisting of finitely nonzero sequences.
The semigroup $e^{\alpha B_c + \beta B_r}$ will not, however,
typically have any nontrivial invariant subspace. To satisfy the
requirements of the Finite Controllability Theorem
one never uses a linear combination of the operators.
Further, as we have seen, the key to controllability is that each
operator has a different invariant subspace within the set of
finite superpositions.

\subsubsection{Explicit finite controllability scheme}
The property that both control vector fields are never used simultaneously is
exploited by Law and Eberly~\cite{LawEberly} and Kneer and
Law~\cite{KneerLaw} in order to devise a explicit scheme for the production
of a finite superposition of eigenstates from another finite
superposition in the control of a spin-half particle coupled to a
harmonic oscillator (in the Lamb-Dicke limit). It shows that if $x$ can be transferred to
$y$ by a series of such ``single nonzero $u_i$" moves then the
transfer from $y$ to $x$ is also possible.

Specifically the Law-Eberly scheme~\cite{LawEberly} to transfer
any eigenstate $|i\rangle$ to any other eigenstate $|j\rangle$
involves the alternate use of transitions generated by spin
reversal ($\pi$-pulses of $E_c$) and transitions generated by
$\pi$-pulses of $E_r$ which convert from a state in which the
oscillator has energy $E_i$ and spin down to a state in which the
energy of the oscillator in altered by one unit and the spin is
flipped as well (see equation (\ref{eq:controlops})).  For example
suppose we wish to drive a state from the $|\downarrow,n\rangle$
to $|\uparrow,n-2\rangle$ (see Fig.~\ref{fig:spin2HO}). This can
be done using $B_r$ to drive the system from
$|\downarrow,n\rangle$ to $|\uparrow,n-1\rangle$, $B_c$ to drive
the system from $|\uparrow,n-1\rangle$ to $|\downarrow,n-1\rangle$
and finally $B_r$ to go from $|\downarrow,n-1\rangle$ to
$|\uparrow,n-2\rangle$.

We note that this scheme works both in the Lamb-Dicke limit and beyond the Lamb-Dicke limit.  
In the Law-Eberly scheme, the $\pi$-pulses of $E_c$ are all of the same time
duration because in the Lamb-Dicke limit, all the carrier
transitions are equally strong. However, the coupling strengths of
the red-sideband transitions are proportional to $\sqrt{n}$, and
therefore the $\pi$-pulses of $E_r$ are shorter in duration as
eigenstates of higher $n$ are addressed.  In order to generate an
arbitrary superposition of a finite number of eigenstates,
starting from another arbitrary superposition, an additional trick
is to go through the ground state of the system which acts as a
``pass state''~\cite{TuriniciReview2000}.  It is possible to
provide an explicit algorithm which will drive the system from any
finite superposition to any other finite superposition.

To prepare an arbitrary finite superposition, the simplest path is
to take the system through the ground state. One assumes that the
desired state is the initial state and then designs a sequence of
alternating pulses of the $E_c$ and $E_r$ fields that would take
this state to the ground state
$|\downarrow,0\rangle$~\cite{KneerLaw}. The actual sequence that produces the superposition is the
time-reversed sequence that was designed. For example, if the
desired superposition is
$(|\uparrow,3\rangle+|\downarrow,2\rangle)/\sqrt 2$, the sequence
of pulses that will transfer this state to the ground state is
$ E_c^{(1)}(\pi)\ $$E_r^{(2)}(\phi_2)\ E_c^{(3)}(\phi_3)\
E_r^{(4)}(\phi_4)\ E_c^{(5)}(\phi_5)\ E_r^{(6)}(\phi_6)\ E_c^{(7)}(\phi_7).$
The action of each pulse is the following: $E_c^{(1)}$ is a $\pi$
pulse of the carrier field that moves the state
$|\uparrow,3\rangle$ to $|\downarrow,3\rangle$. (Simultaneously,
the population in $|\downarrow,2\rangle$ is transferred to
$|\uparrow,2\rangle$).  $E_r^{(2)}$ is a pulse of the red-sideband
field that moves between the states $|\downarrow,3\rangle$ and
$|\uparrow,2\rangle$. Since there is already a superposition of
the two states, the duration of the red-sideband field is shorter
than that of a $\pi$-pulse.  Simultaneously, a superposition of
$|\downarrow,2\rangle$ and $|\uparrow,1\rangle$ is created.  The
next transition $E_c^{(3)}(\phi_3)$ transfers population between
$|\uparrow,2\rangle$ and $|\downarrow,2\rangle$, and again is
shorter than a $\pi$ pulse.  This sequence progresses till all the
population is in $|\downarrow,0\rangle$.  The actual sequence is
the time-reversed sequence of the one that is described above ---
this creates the desired superposition from the initial ground
state.

If one were to transfer an arbitrary initial superposition to an
arbitrary final superposition of eigenstates, one employs the
above algorithm twice.  The sequences A and B that take the system
from the initial and final superpositions respectively to the
ground state are first calculated.  Then the sequence A is first
applied taking all the population to the ground state.  The time
time-reversed sequence of B is then applied which takes the
population to the desired final superposition.  
Clearly, this scheme works in
finite time only if the initial and final states are both
superpositions of a {\em finite} number of states.

Note that finite superpositions are dense in the Hilbert space of
all possible states. Hence from our Lie algebra analysis and the
use of the Law-Eberly algorithm we have
\begin{prop}
The span of the Lie algebra generated by the operators $B_c$ and
$B_r$ for the quantum control system in Eq.~(\ref{controlsystem})
is infinite-dimensional and the reachable set, which is dense in
the Hilbert space of all states, includes all finite
superpositions.
\end{prop}
Note that this provides an explicit dense subspace controllability
result which is hard to prove by abstract methods (see
\cite{BMS1982} and \cite{Z2000}).
Note also that the proof of controllability that
Law and Eberly give of what they term ``arbitrary control" might
be more accurately described as demonstrating that any state in
$l_0$ can be mapped to any other state in $l_0$, staying within
$l_0$ (see the discussion in subsection~\ref{discussion}).

\subsubsection{Red and blue sideband controlled trapped-ion qubit}
The importance of the requirement of the Finite Controllability Theorem that the the system is unit vector controllable on $\mathcal H_1$ is seen by examining the trapped-ion qubit driven by the red and blue sideband fields alone. Recall that  a monochromatic field of frequency $\omega=\omega_0-\omega_m$  ($\omega=\omega_0+\omega_m$) causes resonant transitions between
states $|\downarrow,n\rangle$ and $|\uparrow,n-1\rangle$ ($|\downarrow,n\rangle$ and $|\uparrow,n+1\rangle$), i.e., produces red (blue) sideband transitions. The Schr\"{o}dinger equation in the interaction picture is written as
\begin{eqnarray}
\dot{Y} & = & (v_1(t) B_r + v_2(t) B_b)Y,
\end{eqnarray}
where $v_1(t)$ and $v_2(t)$ are of the form of $v(t)$ defined before. Then in terms of the Laguerre polynomials of the first order and with argument $\eta^2$, 
\begin{eqnarray}
B_r & = & \left(\begin{array}{c|cc|cc|cc}
0 & 0 & 0 & 0 & 0 & 0 & \ldots \\
\hline
0 & 0 & L_0^{(1)} & 0 & 0 & 0 & \ldots \\
0 & -L_0^{(1)} & 0 & 0 & 0 & 0 & \ldots \\
\hline
0 & 0 & 0 & 0 & L_1^{(1)} & 0 & \ldots \\
0 & 0 & 0 & -L_1^{(1)} & 0 & 0 & \ldots \\
\hline
0 & 0 & 0 & 0 & 0 & 0 & \ldots \\
\vdots & \vdots & \vdots & \vdots & \vdots & \vdots & \ddots \\
\end{array} \right). \nonumber
\end{eqnarray}
\begin{eqnarray}
B_b & = &  \left(\begin{array}{cc|cc|cc|c}
0 & 0 & 0 & L_0^{(1)}  & 0 & 0 & \ldots \\
0 & 0 & 0 & 0 & 0 & 0 & \ldots \\
\hline
0 & 0 & 0 & 0 & 0 & L_1^{(1)} & \ldots \\
-L_0^{(1)}  & 0 &  & 0 & 0 & 0 & \ldots \\
\hline
0 & 0 & 0 & 0 & 0 &  & \ldots \\
0 & 0 & -L_1^{(1)} & 0 &  & 0 & \ldots \\
\hline
\vdots & \vdots & \vdots & \vdots & \vdots & \vdots & \ddots \\
\end{array} \right). \nonumber
\end{eqnarray}

\begin{figure}[h!]
\centering
\rotatebox{0}{\includegraphics[width=3in]{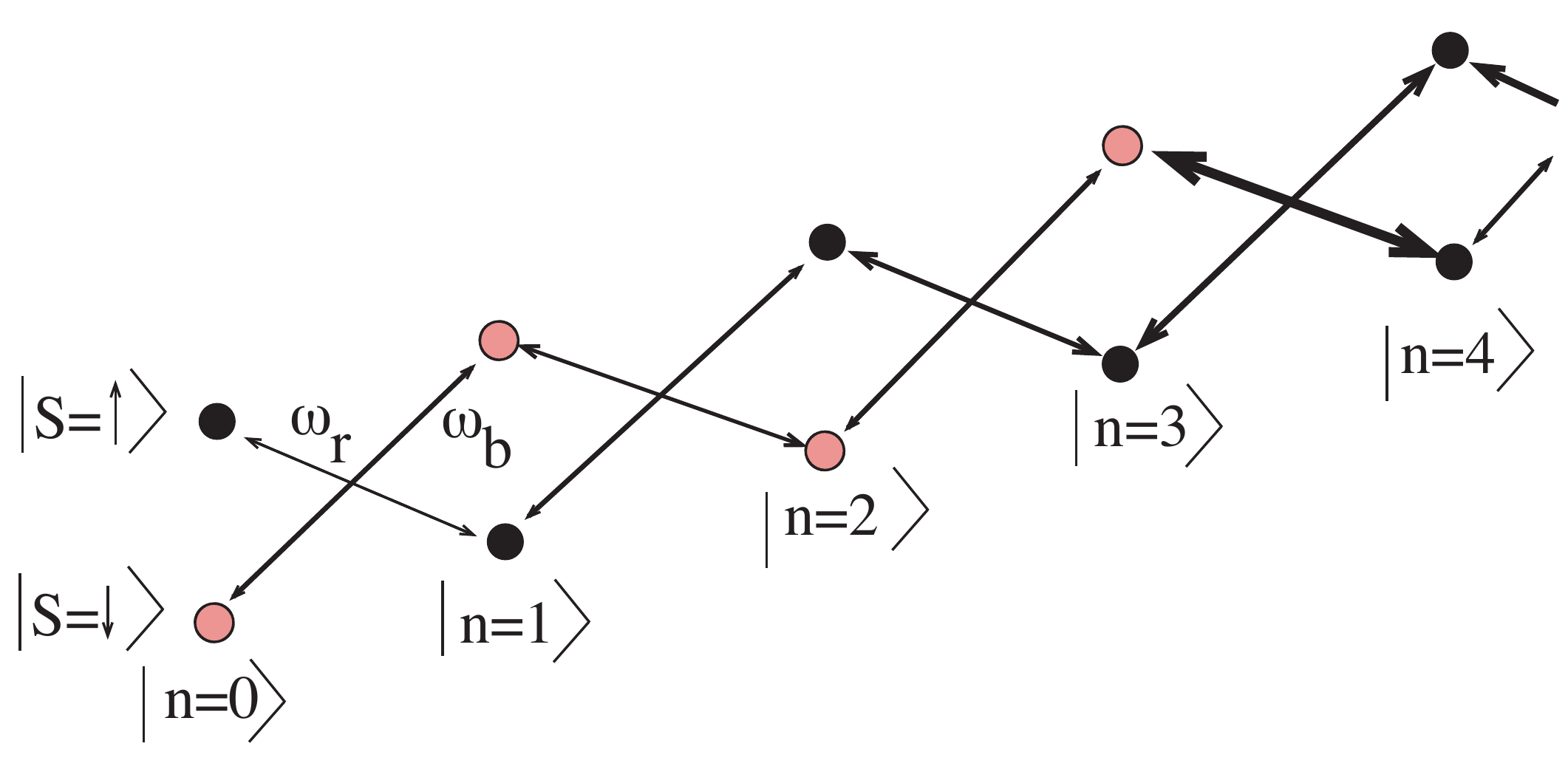}}
\caption{Graphical representation of the coupled spin-half quantum
harmonic oscillator system driven by sinusoidal resonant fields
$\omega_r$ and $\omega_b$ as shown. The strengths of the
$\omega_r$ and $\omega_b$ transition couplings increase as the square root of
$n$.\label{fig:spin2HOrb} }
\end{figure}

From Fig.~\ref{fig:spin2HOrb}, it is seen that the sequentially
connected eigenstates can be looked at as an infinite set of
finite-dimensional subspaces with the space consisting of
$|\downarrow,0\rangle$ and $|\uparrow,0\rangle$ being equal to $\mathcal H_1$.  Further, when
operators $B_r$ and $B_b$ are applied \emph{sequentially}, each subspace
$\mathcal H_i$ can be transferred to $\mathcal H_{i-1}$.  However, by sequential application of
the two operators, any finite superposition of eigenstates cannot be
transferred to the ground state in finite time.  This is because the subspace $\mathcal H_1$ is not transitively connected and therefore not unit vector controllable.  

Note that this is a special case because these two control fields split the Hilbert space into two unconnected subspaces (denoted in the graph by the states with the grey nodes and the black nodes).  In each of these subspaces, the Finite Controllability Theorem holds.  That is, one might transfer finite superpositions to other finite superpositions in one subspace or the other.

\subsection{System 3: Control of an N-level ion trapped in a
quadratic potential} We note that the above example may be extended to
models of an $N$-level system coupled to a quantum harmonic
oscillator. Without loss of generality, it can be assumed that the
energy levels in the $N$-level system are not equally spaced (for example atomic
levels). If $N-1$ monochromatic, resonant fields are available to
couple every pair of adjacent energy levels, the $N$-level system
itself is transitively connected.  It is necessary to have one
more control field in order to make the $\mathcal{H}_i$ to
$\mathcal{H}_{i-1}$ transition (ladder transition).  There are
multiple control schemes that are in keeping with the spirit
behind the Finite Controllability Theorem. Consider
the specific case when $N=3$ (the generalization to higher $N$ is
fairly obvious).  The eigenstates of the coupled system are shown
graphically below in Fig.~\ref{fig:spin3HO1}(a). The resonant fields
$\omega_{c1}$ and $\omega_{c2}$ that transitively connect the
eigenstates of the ion are also shown.

\begin{figure}[h!]
\centering
\rotatebox{0}{\includegraphics[width=3in]{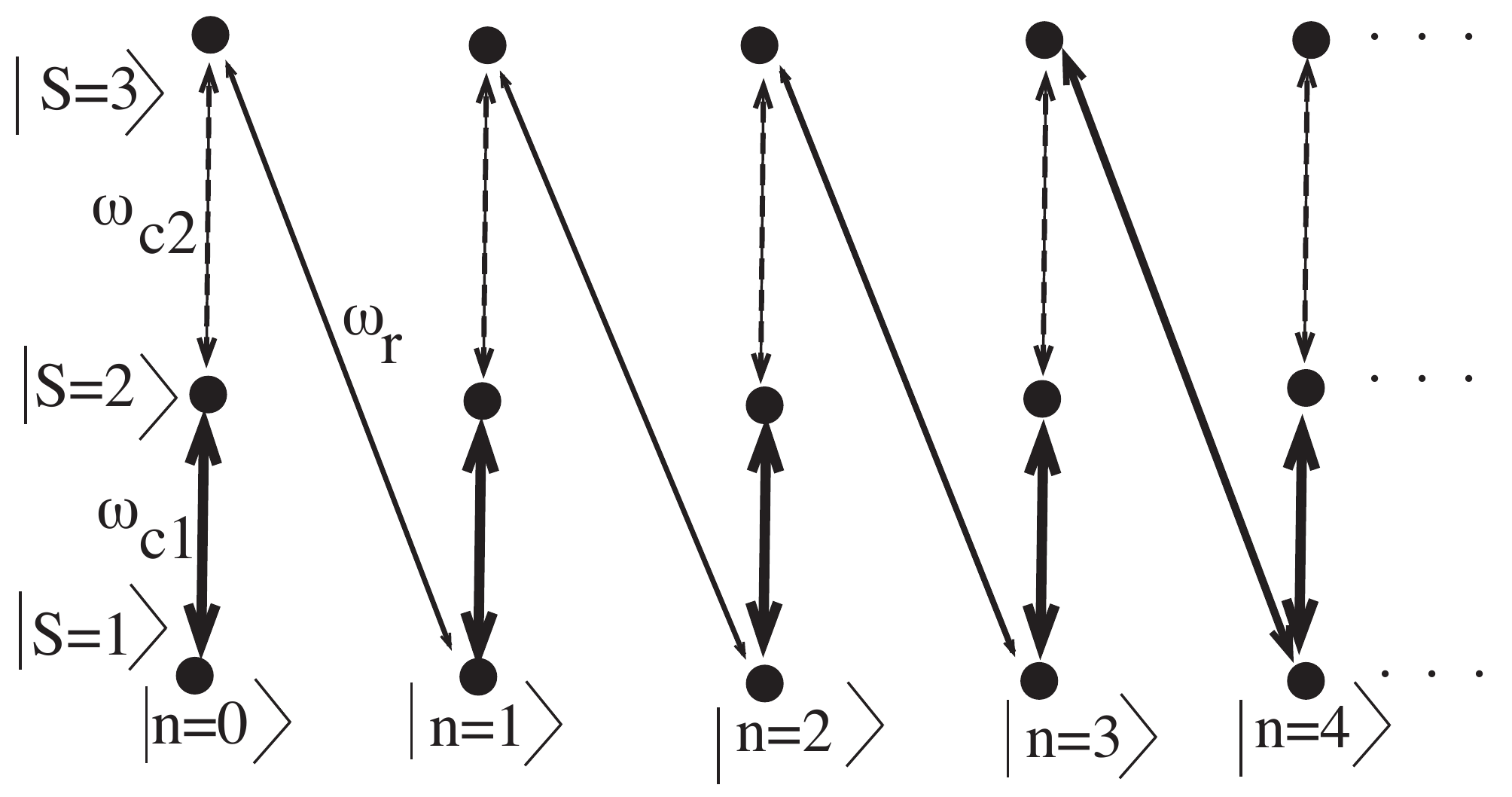}}\\
\rotatebox{0}{\includegraphics[width=3in]{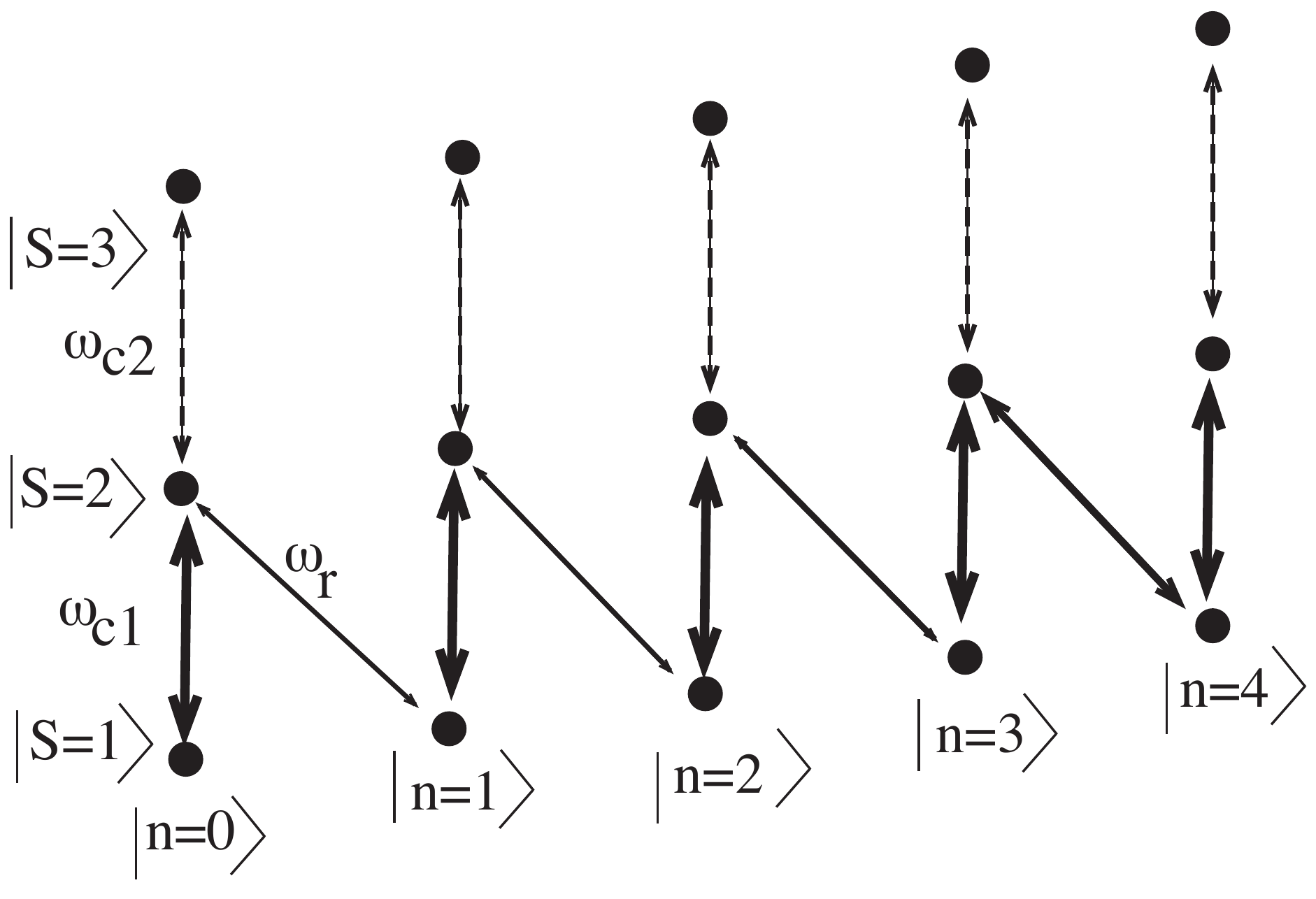}}
\caption{(a)Top: Graphical representation of the N-level ion coupled to
quantum harmonic oscillator driven by sinusoidal resonant fields
$\omega_{c1}$, $\omega_{c2}$ and $\omega_r$ as shown. The
strengths of the $\omega_{c1}$ and $\omega_{c2}$ transition
couplings are independent of the harmonic oscillator quantum
number $n$, whereas the strengths of the $\omega_r$ transition
couplings depend on $n$. (b) Bottom: Graphical representation of the N-level ion coupled to
quantum harmonic oscillator driven by sinusoidal resonant fields
$\omega_{c1}$, $\omega_{c2}$ and $\omega_r$ as shown. This control
scheme, although not very intuitive or elegant, is based on the
Finite Controllability Theorem. \label{fig:spin3HO1} }
\end{figure}

Consider an additional field that accomplishes the ladder
transition by connecting the $|1,n\rangle$ and the $|N,n-1\rangle$
states.  In this case, the eigenstates of the coupled system are
sequentially connected, and $\l_0$ subspace control proceeds
exactly as in the previous example.  The control
matrices also look very similar. As before, in the interaction
picture, the Schr\"{o}dinger equation written as
\begin{eqnarray}
\dot{Y} & = & (u_1(t) B_{c1} +u_2(t) B_{c2} + v(t) B_r)Y,
\end{eqnarray}
where $u_{1,2}(t)$ and $v(t)$ are defined as before. Then
qualitatively, with `X', `Y', `$Z_i$'s denoting non-zero, real matrix
elements,
\begin{eqnarray}
B_{c_1} & = & i \left(\begin{array}{ccc|ccc|cc}
0 & X & 0 & 0 & 0 & 0 & 0 & \ldots \\
X & 0 & 0 & 0 & 0 & 0 & 0 & \ldots \\
0 & 0 & 0 & 0 & 0 & 0 & 0 & \ldots \\
\hline
0 & 0 & 0 & 0 & X & 0 & 0 & \ldots \\
0 & 0 & 0 & X & 0 & 0 & 0 & \ldots \\
0 & 0 & 0 & 0 & 0 & 0 & 0 & \ldots \\
\hline
0 & 0 & 0 & 0 & 0 & 0 & 0 & \ldots \\
\vdots & \vdots & \vdots & \vdots & \vdots & \vdots & \vdots & \ddots \\
\end{array} \right).
\end{eqnarray}
\begin{eqnarray}
B_{c_2} & = & i \left(\begin{array}{ccc|ccc|cc}
0 & 0 & 0 & 0 & 0 & 0 & 0 & \ldots \\
0 & 0 & Y & 0 & 0 & 0 & 0 & \ldots \\
0 & Y & 0 & 0 & 0 & 0 & 0 & \ldots \\
\hline
0 & 0 & 0 & 0 & 0 & 0 & 0 & \ldots \\
0 & 0 & 0 & 0 & 0 & Y & 0 & \ldots \\
0 & 0 & 0 & 0 & Y & 0 & 0 & \ldots \\
\hline
0 & 0 & 0 & 0 & 0 & 0 & 0 & \ldots \\
\vdots & \vdots & \vdots & \vdots & \vdots & \vdots & \vdots & \ddots \\
\end{array} \right).
\end{eqnarray}
\begin{eqnarray}
B_r & = & \left(\begin{array}{ccc|ccc|cc}
0 & 0 & 0 & 0 & 0 & 0 & 0 & \ldots \\
0 & 0 & 0 & 0 & 0 & 0 & 0 & \ldots \\
0 & 0 & 0 & Z_1 & 0 & 0 & 0 & \ldots \\
\hline
0 & 0 & -Z_1 & 0 & 0 & 0 & 0 & \ldots \\
0 & 0 & 0 & 0 & 0 & 0 & 0 & \ldots \\
0 & 0 & 0 & 0 & 0 & 0 & Z_2 & \ldots \\
\hline
0 & 0 & 0 & 0 & 0 & -Z_2 & 0 & \ldots \\
\vdots & \vdots & \vdots & \vdots & \vdots & \vdots & \vdots & \ddots \\
\end{array} \right).
\end{eqnarray}

Another control scheme that exemplifies this theorem is an
additional field that accomplishes the ladder transition by
connecting the $|2,n\rangle$ and the $|N,n-1\rangle$ states, as
shown in Fig.~\ref{fig:spin3HO1}(b). Control in this case is a little
more complicated, because the fields that transitively connect the
$N$-level system must be correctly applied in order to bring the
population to the $|2,n\rangle$ states before applying the ladder
transition.  The control matrices do not look as elegant as those
in the first case, and practically this is a weaker scheme.  As
$N$ increases, it is clear that the number of such schemes
increases. However, the sequentially connected scheme is the one
that is easiest to implement.  The Law-Eberly and Kneer-Law schemes can be modified to provide explicit control schemes in both these cases.

\subsection{System 4: Spin-half particle coupled to two harmonic
oscillators}

In this section we consider another paradigm of quantum computing - the trapped-electron qubit, which can be well-modelled as a
spin-half particle coupled to two harmonic oscillators.  We show that this is a system that is less controllable than the spin-half particle coupled to one harmonic oscillator.  In this system, it is not 
possible to make a transfer from any finite superposition of states
to any other finite superposition. This implies in particular that 
the system cannot be finitely controllable. As we illustrate below, 
the natural control sequence does not allow one to remain within any one 
of the nested invariant subspaces spanned by the 
eigenstates. In addition the nested sequence 
of invariant subspaces is infinite-dimensional.

\begin{figure}[h!]
\centering
\includegraphics[width=3in]{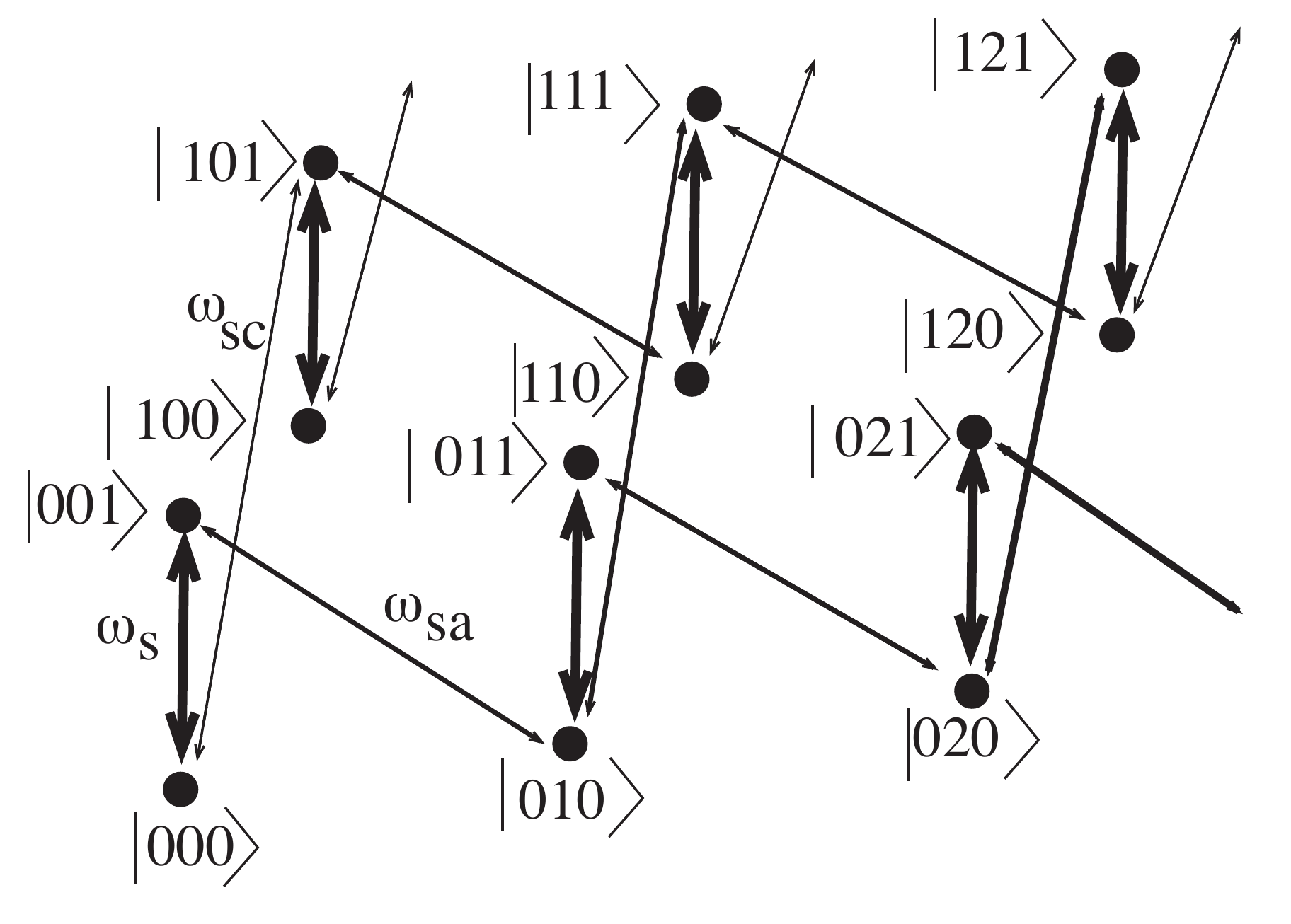}
\caption{ Transfer graph of the controlled
trapped-electron. States are denoted $|nlj\rangle$, where $n$ is the cyclotron harmonic oscillator number state, $l$ is the axial harmonic oscillator
 number state, and $j$ is
the spin state.
Lines marked $\omega_s$, $\omega_{sa}$, and $\omega_{sc}$ indicate spin, spin-axial
and spin-cyclotron transitions, respectively.\label{fig:spin2HOHO}}
\end{figure}

As detailed in Ref.~\cite{PedersenQIP2008}, the energy eigenstates of a spin-half particle coupled to two harmonic oscillators (called the cyclotron oscillator and axial oscillator respectively) can be written as $|nlj\rangle$ where $n$
refers to the number state of the cyclotron oscillator, $l$ refers
to the number state of the axial oscillator, and $j$ refers to the spin state (up or down).  The system is addressed by three control fields.  The spin qubit is controlled using a 
field of angular frequency $\omega_s$ that connects states $|\ n\ l\downarrow\rangle$ and $|n\ l\ \uparrow\rangle$.  The spin-axial transition is controlled by 
field of angular frequency $\omega_{sa}$ that connects states $|n\ l\ \downarrow\rangle$ and $|n\ l+1\ \uparrow\rangle$.  A spin-cyclotron transition is controlled by a field of angular frequency $\omega_{sc}$  that connects states $|n\ l\ \downarrow\rangle$ and $|n-1\ l\ \uparrow\rangle$.  The transfer graph in Fig.\ref{fig:spin2HOHO} clearly indicates that the three fields
transitively connect all the eigenstates of the
spin-axial-cyclotron system.

As in the case of the trapped-ion, the evolution equation is clearer in the interaction picture                     and is derived in Appendix~\ref{appspinHO}. 
Writing down the control matrices explicitly is not trivial, and doesn't provide significant insights into the problem.  Instead, one can look at the structure of these matrices as below.  The control matrices can be written in the basis of trapped-electron eigenstates denoted by $|nlj\rangle$, where $n$ the cyclotron state, $l$ the axial state, and $j$ is the spin state.
Qualitatively, S, A and C indicate spin, spin-axial
and spin-cyclotron transitions, respectively, and the prefactors
indicate the relative strengths.  
The eigenstates are ordered as: $|l=0,n=0,\downarrow\rangle, \ |l=0,n=0,\uparrow\rangle, \ |l=0,n=1,\downarrow\rangle, \ |l=0,n=1,\uparrow\rangle, \ |l=0,n=2,\downarrow\rangle, \ |l=0,n=2,\uparrow\rangle, \ \ldots, |l=1,n=0,\downarrow\rangle, \ |l=1,n=0,\uparrow\rangle, \ |l=1,n=1,\downarrow\rangle, \ |l=1,n=1,\uparrow\rangle, \  |l=1,n=2,\downarrow\rangle, \ |l=1,n=2,\uparrow\rangle, \ \ldots. $

The control matrices that describes the spin-flip and spin-axial
transitions have the form\begin{eqnarray}
B_s  =  i \left[\begin{array}{c||c||c||c}
X_S & 0 & 0 & \ldots \\
\hline
\hline
0 & X_S & 0 & \ldots \\
\hline
\hline
0 & 0 & X_S & \ldots \\
\hline
\hline
\vdots & \vdots & \vdots & \ddots \\
\end{array} \right],\\
B_{sa}  =  i \left[\begin{array}{c||c||c||c}
X_A & 0 & 0 & \ldots \\
\hline
\hline
0 & X_A & 0 & \ldots \\
\hline
\hline
0 & 0 & X_A & \ldots \\
\hline
\hline
\vdots & \vdots & \vdots & \ddots \\
\end{array} \right],
\end{eqnarray}
\noindent respectively. 

\noindent Here $X_S$ and $X_A$ are the  infinite block matrices
\begin{eqnarray}
X_S  =  \left[\begin{array}{cc|cc|cc|c}
0 & S & 0 & 0 & 0 & 0 & \ldots  \\
S & 0 & 0 & 0 & 0 & 0 & \ldots \\
\hline
0 & 0 & 0 & S & 0 & 0 & \ldots \\
0 & 0 & S & 0 & 0 & 0 & \ldots \\
\hline
0 & 0 & 0 & 0 & 0 & S & \ldots \\
0 & 0 & 0 & 0 & S & 0 & \ldots  \\
\hline
\vdots & \vdots & \vdots & \vdots & \vdots & \vdots & \ddots \\
\end{array} \right],\\
X_A  =   \left[\begin{array}{cc|cc|cc|c}
0 & 0 & 0 & 0 & 0 & 0 & \ldots  \\
0 & 0 & A & 0 & 0 & 0 & \ldots \\
\hline
0 & A & 0 & 0 & 0 & 0 & \ldots \\
0 & 0 & 0 & 0 & A\sqrt 2 & 0 & \ldots \\
\hline
0 & 0 & 0 & A\sqrt 2 & 0 & 0 & \ldots \\
0 & 0 & 0 & 0 & 0 & 0 & \ldots  \\
\hline
\vdots & \vdots & \vdots & \vdots & \vdots & \vdots & \ddots \\
\end{array} \right].
\end{eqnarray}
\noindent respectively. 

The control matrix describing the spin-cyclotron transition has the form\begin{eqnarray}
B_{sc}  =  i \left[\begin{array}{c||c||c||c}
0 & X_C & 0 & \ldots \\
\hline
\hline
X_C^{\dag} & 0 & X_C & \ldots \\
\hline
\hline
0 & X_C^{\dag} & 0 & \ldots \\
\hline
\hline
\vdots & \vdots & \vdots & \ddots \\
\end{array} \right],
\end{eqnarray}
\noindent where $X_C$ is an infinite block matrix of the form
\begin{eqnarray}
X_C  =  \left[\begin{array}{cc|cc|cc|c}
0 & C & 0 & 0 & 0 & 0 & \ldots  \\
0 & 0 & 0 & 0 & 0 & 0 & \ldots \\
\hline
0 & 0 & 0 & C & 0 & 0 & \ldots \\
0 & 0 & 0 & 0 & 0 & 0 & \ldots \\
\hline
0 & 0 & 0 & 0 & 0 & C\sqrt 2 & \ldots \\
0 & 0 & 0 & 0 & 0 & 0 & \ldots  \\
\hline
\vdots & \vdots & \vdots & \vdots & \vdots & \vdots & \ddots \\
\end{array} \right].
\end{eqnarray}

When the three transitions are
applied sequentially, it is possible to transfer population from
subspace $\mathcal H_i$ to subspace $\mathcal H_{i-1}$ only under certain conditions.  Trivially, if one wishes to transfer population within one of the harmonic oscillator states, the problem is the same as that of the trapped ion.  Also, it is
easy to see that the levels are connected in such a way that the
system is eigenstate controllable in the sense that the population can be
coherently transferred from any eigenstate to any other
eigenstate. For example, consider the set of eigenstates
illustrated in Fig.\ref{fig:spin2HOHO}. The condition for eigenstate
controllability is that the pulses of frequency $\omega_{sa}$,
$\omega_s$ and $\omega_{sc}$ must be applied {\it sequentially},
and not simultaneously. For example, let us say we want to
transfer the $|000\rangle$ state to the $|111\rangle$ state. We
can do so by the pulse sequence: $p_s(\pi)$ that transfers to $|001\rangle$,
$p_{sa}(\pi)$ that transfers to $|010\rangle$, and then $p_{sc}(\pi)$ that transfers to $|111\rangle$.

Superficially this system appears to be very similar to the trapped-ion
system.  However, it is seen that this system is not finitely controllable; that is, even though the system is eigenstate
controllable, it is not possible to transfer a superposition of eigenstates to
the ground state even with sequential applications of the field.
This can be seen visually from the control graph in Fig.\ref{fig:spin2HOHO}, or by examining
the control matrices as follows.

For example, let us say we want to
transfer the $|000\rangle + |111\rangle$ state to the $|000\rangle$ state. The pulse sequence: $p_s(\pi)$, $p_{sa}(\pi)$, $p_{sc}(\pi)$ that transfers the $|000\rangle$ state to the $|111\rangle$ state also transfers the $|111\rangle$ state to the $|000\rangle$ state.  Explicitly, $p_s(\pi)$ transfers $|111\rangle$ to $|110\rangle$, but also transfers  $|000\rangle$ to $|001\rangle$;  $p_{sa}(\pi)$ transfers $|110\rangle$ to $|101\rangle$, but also transfers  $|001\rangle$ to $|010\rangle$; and $p_{sc}(\pi)$ transfers $|101\rangle$ to $|000\rangle$, but also transfers  $|010\rangle$ to $|111\rangle$.  The pulse sequence that transfers population down one of the harmonic oscillator ladders, transfers population up the other harmonic oscillator ladder.  Thus we cannot apply 
the Finite Controllability Theorem using the above basis.

This lack of controllability can be attributed to the
fact that the invariant subspaces are not finite-dimensional.  In fact this
system presents an infinite number of nested \emph{infinite}
-dimensional subspaces.

Thus we see that eigenstate controllability is a much weaker condition than
subspace controllability. Simply having vector fields that sequentially
connect the eigenstates is sufficient for 
eigenstate controllability.  This is in contrast to the controllability of finite-dimensional quantum systems where eigenstate controllability implies controllability of finite superpositions.

This analysis provides important input into the the development of quantum gates in the trapped-electron system --- it is not possible to use Law-Eberly type methods to construct arbitrary quantum gates in this system.

\section{Summary}
Many of the novel questions that arise in laying the ground work
for quantum computing can be thought of as questions about the
controllability of Schr\"odinger's equation.
In this paper we prove a Finite Controllability Theorem which is useful for analyzing certain infinite-dimensional quantum control problems. We discuss several related physical systems that are models for quantum computing.

Among the more interesting paradigms of quantum computing is the trapped ion
modeled as a spin-half particle coupled to a quantum harmonic
oscillator. In this paper, we discuss a general setting for this
type of problem based on infinite-dimensional differential
equations and Lie groups acting on a Hilbert space.  This allows
us to explore the Lie algebraic approach to controllability in
this setting. In particular, we show that even though the formal
Lie algebra associated with the Jaynes-Cummings model is
infinite-dimensional, explicit finite controllability
results can be determined.  We also establish generalized
controllability criteria for improving the controllability of some important
infinite-dimensional quantum systems.

For the specific quantum computing system of the trapped-ion qubit, we showed that the Law-Eberly control scheme does not require the system to be in the Lamb-Dicke limit.  We also showed that finite controllability cannot be achieved in the trapped-electron quantum computing paradigm, thus limiting the type of quantum operations (gates) that can be executed in this system. 

\bibliographystyle{unsrt}
\bibliography{infinite23}

\appendix
\label{appspinHO}
\noindent {\bf Derivation of the evolution equation for the trapped-ion qubit system}

The trapped-ion qubit is modeled as a spin-half particle coupled to a quantized harmonic oscillator.  Since various aspects of this model are discussed in several
publications, we summarize only the essential features here.  Consider
the description of a particle with two spin states in a quadratic
potential field. The Hamiltonian now includes terms that reflect
both the linear momentum of the particle in the potential field,
and the spin angular momentum of the particle. The Schr\"odinger
equation can be written as a two-component vector equation \beq
\left[ \begin{array}{c} \frac {\partial \psi _+}{\partial t }
\\
\frac {\partial \psi _-}{\partial t } \end{array} \right] & = &
\frac{i}{2}\left[ \begin{array}{cc} \Omega_m-\omega_0 & 0 \\ 0 & \Omega_m
+\omega_0\end{array} \right] \left[
\begin{array}{c} \psi _+ \\ \psi _-
\end{array} \right]\label{eq:FldFreeCoupledSys}
\eeq \noindent where $\Omega_m = \omega_m\left( \frac{\partial ^2 
}{\partial x^2}-x^2 \right)$.  The subscripts $+$ and $-$ refer to the two levels of the
atom (modelled in physics literature as the spin-up and spin-down
states of a spin-$1/2$ particle).  An applied field causes
transitions between the eigenstates of the coupled spin-oscillator
system.

In order to understand the control of this system, we briefly present the a spin-$0$ particle in a quadratic potential with resonant frequency $\omega_m$ being excited by a
traveling wave of central frequency $\omega$ with controllable
amplitude $E(t)$.  This system might be described by an equation of the form \beq
\frac{\partial \psi }{\partial t} = \frac{i\omega_m}{2}\left(
\frac{\partial ^2 }{\partial x^2}-x^2 \right) \psi -
i E(t) \mu \cos (kx-\omega t)\psi ,\label{eq:drivenSHO} \eeq where $\mu$ is a dipole operator that describes the transition couplings between various eigenstates of the system, and $k$ is the wavevector of the applied field.
Let us introduce the differential operators \beq a =
\frac{1}{\sqrt{2}}\left(x+\frac{\partial}{\partial x}\right),\eeq
and \beq a^{\dagger} =
\frac{1}{\sqrt{2}}\left(x-\frac{\partial}{\partial x}\right).\eeq
These are the familiar creation and annihilation operators of the
quantized harmonic oscillator~\cite{Schiffbook}. Introducing the
differential operators corresponding the position of the particle
in the harmonic potential, and using the notation established in the previous subsection, Eq.~\ref{eq:drivenSHO} becomes \beq
\frac{\partial \psi }{\partial t}  &=& -i \omega_m \left(
N+\frac{1}{2} \right) \psi  \nonumber \\
&&- i \left( E(t)\mu \cos (\omega t- kx_0(a+a^{\dagger}) \right)\psi. \eeq \noindent The
product of $k$, the wavelength of the light, and $x_0$, the
amplitude of the zero-point motion of the particle in the harmonic
potential (or the spatial extent of the ground state harmonic
oscillator wave function) is the Lamb-Dicke
parameter $\eta$.  Recall that the driving field is not resonant with the characteristic frequency of the harmonic potential.  Therefore the above equation can be simplified only with the knowledge of the driving frequency, and the matrix elements of the operator $\mu$ that can be coupled by that driving frequency due to energy conservation.

Returning to the system described by
Eq.~\ref{eq:FldFreeCoupledSys}, an applied field causes
transitions between the eigenstates of the coupled spin-oscillator
system.  
The amplitudes corresponding to the fields that cause the
carrier and red transitions are dubbed $E_c$ and $E_r$
respectively.  When both fields are applied simultaneously, ignoring overall additive constants, the control equations take on the form
\beq
\left[ \begin{array}{c} \frac {\partial \psi _+}{\partial t }
\\
\frac {\partial \psi _-}{\partial t } \end{array} \right]  = 
\frac{i}{2}\left[ \begin{array}{cc} \Omega_m -\omega_0 & 0 \\ 0 & \Omega_m 
+\omega_0\end{array} \right] \left[
\begin{array}{c} \psi _+ \\ \psi _-
\end{array} \right] \\ \nonumber
    - i \left[ \begin{array}{cc} 0 & \mu_c E_c(t)+\mu_r E_r(t) \\ \mu_c E_c(t)+\mu_r E_r(t)  & 0\end{array} \right] \left[
\begin{array}{c} \psi _+ \\ \psi _-
\end{array} \right];
\eeq \noindent where, $\mu_{c,r}$ is short for $\mu \cos(\omega_{c,r} t - \eta(a+a^{\dagger}))$.
Here $\mu$ is the dipole operator that describe the strengths of the transitions between the two-level system eigenstates coupled by the fields.  Examining the cosine terms in the off-diagonals, these can be expanded as
\beq
\cos(\omega t- \eta(a+a^{\dagger})) = & \frac{1}{2}\left(\exp(i\omega t) \exp(- i\eta(a+a^{\dagger}))) \right. \\ \nonumber &+\left. \exp(-i\omega t )\exp( i\eta(a+a^{\dagger})))\right).
\eeq
Thinking in terms of the field adding energy to the ionic state (two level system), the term $\exp(-i\omega t )$ adds energy $\hbar\omega$ to a state while the term $\exp(i\omega t )$ removes energy $\hbar\omega$ from a state.  For a true two-level system, one cannot add energy to the excited state, nor remove energy from the ground state.  So, keeping only the energy conserving terms (rotating wave approximation), the control equations become
\beq
\left[ \begin{array}{c} \frac {\partial \psi _+}{\partial t }
\\
\frac {\partial \psi _-}{\partial t } \end{array} \right]& = &
\frac{i}{2}\left[ \begin{array}{cc} \Omega_m -\omega_0 & 0 \\ 0 & \Omega_m
+\omega_0\end{array} \right] \left[
\begin{array}{c} \psi _+ \\ \psi _-
\end{array} \right] \\ \nonumber
&   & - i E_c\left[ \begin{array}{cc} 0 & \tilde{\mu}_c\\ \tilde{\mu}_c^*  & 0\end{array} \right] \left[
\begin{array}{c} \psi _+ \\ \psi _-
\end{array} \right] \\ \nonumber
&   & - i E_r\left[ \begin{array}{cc} 0 &\tilde{\mu}_r\\ \tilde{\mu}_r^*  & 0\end{array} \right] \left[
\begin{array}{c} \psi _+ \\ \psi _-
\end{array} \right].
\eeq \noindent where $\tilde{\mu}_{c,r}$ is short for $\mu \exp(-i\omega_{c,r} t + i\eta(a+a^{\dagger}))$.

In order to write the evolution equation in matrix form, the eigenstates are ordered as
$|\uparrow,0\rangle,\ |\uparrow,1\rangle,\ \ldots,
|\downarrow,0\rangle,\ |\downarrow,1\rangle,\ \ldots$.  The drift
matrix of this system can be written in block matrix
form as: \beq -i \left(\begin{array}{c|c}
\frac{\omega_0}{2}+N\omega_m & 0 \\
\hline
0 &  \frac{-\omega_0}{2}+N\omega_m\\
\end{array} \right), \eeq
\noindent where, $N$ is the previously defined number operator.  The control
matrices are not easily expressed in this basis. 

The evolution equation takes on a much simpler and elegant form in the interaction picture.  Also, the transition matrices are described compactly by using the
three Pauli matrices $\sigma _x$, $\sigma _y$, and $\sigma _z$.
For example, the spin-flip transition is described by the operator $\sigma_+ = \sigma_x+i \sigma_y$ and its hermitian conjugate $\sigma_-$. 
In a coordinate system defined by the eigenbasis of the coupled system ($|S,n\rangle$, where $S$ refers to the spin state and $n$ to the harmonic oscillator number state),
a general matrix element of the control Hamiltonian $\langle S^{\prime} n^{\prime}|H_I^{\prime}|S n \rangle$
may be written as~\cite{RanganPRL2004,WinelandNISTreport1998}:
\begin{small}
\begin{eqnarray*}
&   
-i 2 {\rm Re}\left[\langle S^{\prime}|\sigma_+|S\rangle
\langle n^{\prime}|\exp(i(\eta(a+a^{\dag
})))|n \rangle\right]\\ \nonumber
& = 
-i 2 {\rm Re}\left[\langle S^{\prime}|\sigma_+|S\rangle
\exp(-\eta
/2)\sqrt{\frac{n_<!}{n_>!}}\  (i\eta)^{|n^{\prime}-n|}  
L_{n<}^{|n^{\prime}-n|}(\eta^2)\right].\quad \nonumber
\end{eqnarray*}
\end{small}
\noindent The symbol $n_>$ refers to the larger of $n$ and
$n^{\prime}$, and $n_<$ refers to the smaller of $n$ and
$n^{\prime}$.  Here $L_n^{\alpha}(x)$ is the associated Laguerre
polynomial.  When the applied field of frequency $\omega_c$ connects states
$|\downarrow,n\rangle$ and $|\uparrow,n\rangle$ (carrier
transitions), $n^{\prime}=n$, and when the applied field of frequency $\omega_r$ connects
states $|\downarrow,n\rangle$ and $|\uparrow,n-1\rangle$ (red
sideband transitions), $n^{\prime} = n-1$.  The matrix elements
are zero for all other values of $n^{\prime}$.

\noindent {\bf Derivation of the evolution equation for the trapped-electron qubit system}

We consider a trapped-electron qubit which can be well-modelled as a
spin-half particle coupled to two harmonic oscillators.  As
detailed in Ref.~\cite{PedersenQIP2008}, the relevant quantum states are the spin states of the electron (a spin-half particle), and the harmonic oscillator number states of the cyclotron and axial harmonic potentials.  As before, the evolution is described by
\beq  \left[ \begin{array}{c} \frac {\partial \psi _+}{\partial t }
\\
\frac {\partial \psi _-}{\partial t } \end{array} \right] & = &
-\frac{i}{2}\left[ \begin{array}{cc} \Omega_{tot} & 0 \\ 0 & \Omega_{tot}\end{array} \right] \left[
\begin{array}{c} \psi _+ \\ \psi _-
\end{array} \right]\label{eq:FldFreeCoupledSys2};
\eeq  \noindent where $\Omega_{tot}=\omega_c\left( -\frac{\partial ^2 
}{\partial x_c^2}+\beta x_c^2\right) + \omega_z\left(-\frac{\partial ^2 
}{\partial x_z^2} +  x_z^2 \right) +\omega_s$.  Here, $\omega_s$ refers to the angular frequency associated with the spin-flip of the electron.  The parameter $\beta$ indicates that the cyclotron harmonic potential (indicated by subscript $c$) and  the axial harmonic potential (indicated by subscript $z$) are of different strengths.

In the basis of the harmonic oscillator eigenstates, the drift matrix can be written as \beq
\left[ \begin{array}{cc} \omega'_c N_c +  \omega_z N_z +\frac{\omega_s}{2} & 0 \\ 0 & \omega'_c N_c +  \omega_z N_z
-\frac{\omega_s}{2}\end{array} \right], \eeq where the frequencies
$\omega_c$ and $\omega_z$ refer to the frequencies of the
cyclotron and axial harmonic oscillators, and $\omega_s$ refers
to the spin flip transition frequency.  

The energy eigenstates can be written as $|n
lj\rangle$ where $n$
refers to the number state of the cyclotron oscillator, $l$ refers
to the number state of the axial oscillator, and $j$ refers to the spin state (up or down).  The system is addressed by three control fields.  The spin qubit is controlled using a small transverse magnetic
field that connects states $|\ n\ l\downarrow\rangle$ and $|n\ l\ \uparrow\rangle$.  The spin-axial transition is controlled by a travelling magnetic
field that connects states $|n\ l\ \downarrow\rangle$ and $|n\ l+1\ \uparrow\rangle$.  A spin-cyclotron transition is controlled by an oscillating magnetic field with a spatial gradient  that connects states $|n\ l\ \downarrow\rangle$ and $|n-1\ l\ \uparrow\rangle$.  The transfer graph in Fig.\ref{fig:spin2HOHO} clearly indicates that the three fields
transitively connect all the eigenstates of the
spin-axial-cyclotron system.

As in the case of the trapped-ion, the evolution equation is clearer in the interaction picture. The evolution equation is given by
\begin{eqnarray}
\dot{Z} & = & (\Omega_s(t) B_s + \Omega_{sa}(t) B_{sa} + \Omega_{sc}(t) B_{sc})Z.
\end{eqnarray}
Here $\Omega_s(t)$ is the control parameter for the spin flip transition, $\Omega_{sz}(t)$ is the control
parameter for the spin axial transition, and $\Omega_{sc}(t)$ is the control
parameter for the spin cyclotron transition.
The spin-flip transition operator $B_s =- i/2\left( \sigma_+ \exp(-i\phi)+\sigma_- \exp(i\phi)\right)$.  
The spin-axial transition operator \beq
\begin{split}
B_{sa} = &-i \bigl( \sigma_+
\exp(i(\eta(a^\dagger_z e^{i\omega_z t}+a_z e^{-i\omega_z
t})-\Delta \cdot t-\phi)) \\ &+\sigma_- \exp(-i(\eta(a^\dagger_z
e^{i\omega_z t}+a_z e^{-i\omega_z t})-\Delta \cdot t-\phi))\bigr)
\end{split}
\eeq where the characteristic
frequency of the transition is $\omega_{sa}= \omega_s+\omega_z$, the detuning $\Delta = \omega-\omega_s$, $\phi$ is an experimentally variable phase, 
and $\eta$ is the Lamb-Dicke parameter.
Here $a_z$ and $a^{\dagger}_z$ are annhilation and creation operators
respectively associated with the axial harmonic oscillator.
For $\omega = \omega_{sa} $ (when the applied field is at resonance with the spin-axial transition) only the levels $|\!
n l\!\downarrow\rangle$ and $|\! n (l+1)\!\uparrow\rangle$ are connected.
The spin-cyclotron transition operator $B_{sc} = -i(\sigma_+ a_c
e^{-i\phi}+\sigma_- a_c^\dagger e^{i\phi})$, 
where  $\omega_{sc}=\omega_s-\omega'_c$  is the characteristic
frequency associated with
the spin cyclotron transition.
Here $a_c$ and $a^{\dagger}_c$ are annhilation and creation operators
respectively associated with the cyclotron harmonic oscillator.
When the applied field is in resonance with the spin-cyclotron transition $\omega = \omega_{sc}$, only the levels
$|\! n l\!\downarrow\rangle$, $|\! (n-1) l\uparrow\rangle$
are connected.

\end{document}